\documentclass[12pt]{article}

\usepackage{amssymb}
\usepackage{amsmath}
\usepackage{graphics,graphicx}

\newcommand{\singlespace}{\baselineskip=12pt \lineskiplimit=0pt
\lineskip=0pt}
\newcommand{\beq}{\begin{equation}}
\newcommand{\eeq}{\end{equation}}
\newcommand{\lb}{\label}
\newcommand{\beqar}{\begin{eqnarray}}
\newcommand{\eeqar}{\end{eqnarray}}
\newcommand{\bit}{\begin{itemize}}
\newcommand{\eit}{\end{itemize}}
\newcommand{\barr}{\begin{array}}
\newcommand{\earr}{\end{array}}
\def\ds{\displaystyle}

\def\scalp{\mbox{\boldmath$\, \cdot \,$}}
\def\bob{{\, \underline{\overline{\otimes}} \,}}
\newcommand{\derivative}[2]{\frac{\partial #1}{\partial #2}}

\def\vx{{\bf x}}
\def\vy{{\bf y}}
\def\v0{{\bf 0}}

\def\bA{{\bf A}}
\def\bB{{\bf B}}
\def\Id{{\bf I}}
\newcommand{\stress}{\ensuremath{\mbox{\boldmath$\sigma$}}}

\def\devstress{{\bf S}}
\def\bS{{\bf S}}
\def\b0{{\bf 0}}
\def\gradF{{\bf Q}}

\def\Js{\mbox{$J_2$}}
\def\Jt{\mbox{$J_3$}}

\def\HWt{\mbox{$\theta$}}

\def\yf{F}
\def\tr{\mbox{$\mathsf{tr}\,$}}

\def\Sym{\mbox{$\mathsf{Sym}$}}

\def\AM{ Acta mater.\ }
\def\ACISJ{ ACI Structural Journal\ }

\def\AAM{ Arch. Appl. Mech.\ }

\def\IJNAMG{ Int. J. Numer. Anal. Meth. Geomech.\ }
\def\IJF{ Int. J. Fracture\ }
\def\IJP{ Int. J. Plasticity\ }
\def\IJSS{ Int. J. Solids Structures\ }

\def\JAM{ J. Appl. Mech.\ }

\def\JEM{ ASCE J. Engng. Mech.\ }
\def\JMPS{ J. Mech. Phys. Solids\ }

\def\MRC{ Mechanics Research Communications\ }
\def\PRSL{ Proc. R. Soc. Lond.\ }
\def\QAM{ Quart. Appl. Math.\ }

\def\ZAMP{ Z. Angew. Math. Phys. \ }

\begin{document}

\title{Yield criteria for quasibrittle  \\ 
and frictional materials}

\author{\\Davide Bigoni and Andrea Piccolroaz \\
Dipartimento di Ingegneria Meccanica e \\ Strutturale,
Universit\`a di Trento, \\ Via Mesiano 77, I-38050 Trento, Italia\\
email: bigoni@ing.unitn.it}

\date{August 29, 2002}

\maketitle

\begin{abstract}
\noindent
A new yield/damage function is proposed for modelling the inelastic behaviour 
of a broad class of pressure-sensitive, frictional, ductile and brittle-cohesive 
materials. The yield function allows the possibility of describing a transition between 
the shape of a yield surface typical of a class of materials to that typical of another
class of materals. This is a fundamental key to model the behaviour of materials 
which become cohesive during hardening (so that the shape of the yield surface evolves from that typical
of a granular material to that typical of a dense material), or which decrease cohesion due 
to damage accumulation.
The proposed yield function is shown to agree with a variety of experimental data relative to
soil, concrete, rock, metallic and composite powders, metallic foams, porous metals, and polymers. 
The yield function represents a single, convex and smooth surface in stress space
approaching as limit situations well-known criteria and the extreme limits of convexity 
in the deviatoric plane.
The yield function is therefore a ge\-ne\-ra\-li\-za\-tion of several criteria, including
von Mises, Drucker-Prager, Tresca, modified Tresca, Coulomb-Mohr, modified Cam-clay, and
---concerning the deviatoric section--- Rankine and Ottosen.
Convexity of the function is proved by developing two general propositions 
relating convexity of the yield surface to convexity of the corresponding function. 
These propositions are general and therefore may be employed to generate other convex yield functions.
\end{abstract}


\thispagestyle{empty} 

\newpage

\section{Introduction}
\lb{sec: introduction}

Yielding or damage of quasibrittle and frictional materials 
(a collective denomination for soil, concrete, rock, 
granular media, coal, cast iron, ice, porous metals, metallic foams, as well as certain types of ceramic)
is complicated by many effects, including dependence on the first and third stress invariants
(the so-called `pressure-sensitivity' and `Lode-dependence' of yielding), and
represents the subject of an intense research effort. Restricting attention to the 
formulation of yield criteria, 
research moved in two directions: one was to develop such criteria on the basis of
micromechanics considerations, while another was to find direct interpolations to experimental
data. Examples of yield functions generated within the former approach are numerous and,
as a paradigmatic case, we may mention
the celebrated Gurson criterion (Gurson, 1977). The latter approach was also broadly
followed, providing some very successful yield conditions, such as for instance the Ottosen criterion 
for concrete (Ottosen, 1977). Although very fundamental in essence, the micromechanics approach has 
limits however, particularly
when employed for geomaterials. For instance, 
it is usually based on variational formulations, 
possible ---for inelastic materials--- only for solids obeying the postulate of maximum dissipation 
at a microscale, which is typically violated for frictional materials such as for instance soils.

A purely phenomenological point of view is assumed in the present article, wherein a new yield
function\footnote{
We need not distinguish here between yield, damage and failure. Within a phenomenological 
approach, all these situations 
are based on the concept of stress range, bounded by a given hypersurface defined in stress space.}
is formulated, tailored to interpolate experimental results for quasibrittle and frictional materials, 
under the assumption of isotropy. The interest in this proposal lies in the features evidenced by the 
criterion. These are:
\begin{itemize}
\item finite extent of elastic range both in tension and in compression;
\item non-circular deviatoric section of the yield surface, which may approach 
both the upper and lower convexity limits for extreme values of material parameters;
\item smoothness of the yield surface;
\item possibility of stretching the yield surface to extreme shapes 
and related capability of interpolating a broad 
class of experimental data for different materials;
\item reduction to known criteria in limit situations;
\item convexity of the yield function (and thus of the yield surface);
\item simple mathematical expression.
\end{itemize}
None of the above features is {\it essential}, in the sense that a plasticity theory can be
developed without all of the above, but all are {\it desirable} for the development
of certain models of interest, particularly in the field of geomaterials. This is a crucial point, deserving
a carefully explanation. In particular, while some of the above requirements have a self-evident
meaning, smoothness and convexity need some discussion. 

Although experiments are inconclusive in this respect (Naghdi et al. 1958; Paul, 1968; Phillips, 1974),
theoretical speculations (sometimes criticized, Naghdi and Srinivasa, 1994) 
suggest that corners should be expected to form in the yield surface
for single crystals and polycrystals (Hill, 1967). Therefore,
smoothness of the yield surface might be considered a mere simplification in the constitutive modelling of metals.
However, the si\-tuation of quasibrittle and frictional materials is completely different. For such
materials, in fact, evidence supporting corner formation is weak\footnote{Some argument in favour
of corner formation in geomaterials have been given by Rudnicki and Rice (1975).}, so that,
presently, smoothness of the yield surface is a broadly employed concept and models developed under this 
assumption are still very promising. Moreover, corners often are included in the constitutive description 
of a material for the mere fact that an appropriate, smooth yield function is simply not available
(this is usually the case of the apex of the Drucker-Prager yield surface and of the corner which may
exist at the intersection of a smooth, open yield surface with a cap).

Regarding convexity of the yield surface, we note that this follows for polycrystals from Schmid
laws of single crystals (Bishop and Hill, 1951; Mandel, 1966). However, differently from smoothness, 
convexity is supported
by experiments in practically all materials and is a useful mathematical property, which is the 
basis of limit analysis and 
becomes of fundamental importance in setting
variational inequalities for plasticity (Duvaut and Lions, 1976).
We may therefore conclude that ---in the absence of a clear and specific motivation--- 
it is not sensible to employ a yield function that violates convexity.

A number of failure surfaces have been proposed meeting some of the above requirements, 
among others, we quote the 
Willam and Warnke (1975), Ottosen (1977) and Hisieh et al. (1982) criteria
for concrete, the  Argyris et al. (1974), Matsuoka and Nakai (1977),  Lade and Kim (1995), and Lade (1997) 
criteria for soils. For all these criteria, while some
information can be found about the range of parameters corresponding to convexity of 
the yield {\it surfaces}, nothing is known about the convexity of the corresponding 
yield {\it functions}. 

Convexity of a yield function implies convexity of the corresponding 
yield surface, but convexity of a level set of a function does not imply convexity of the function itself.
While it can be pointed out that a convex yield function can {\it in principle} always 
be found to represent a convex yield surface, the `practical problem' of finding it in a reasonably 
simple form may be a formidable one. From this respect, general propositions would be of interest, but the only
contribution of which the authors are aware in this respect is quite recent (Mollica and Srinivasa, 2002). 
A purpose of the present paper is to provide definitive results in this direction. 
In particular, the range of material parameters corresponding to convexity of the 
yield function proposed in this paper
is obtained by developing a general proposition that can be useful for 
analyzing convexity of a broad class of yield functions. The proposition is finally extended 
to introduce the possibility of describing a modification in shape of the deviatoric section with 
pressure. The propositions are shown to be constructive, in the sense that these may be employed to
generate convex yield functions (examples of which are also included).

Beyond the issue of convexity, the central purpose of this paper is 
the proposal of a yield criterion [see eqns. 
(\ref{eq: yield function})-(\ref{gi})]. This meets all of the above-listed requirements and 
can be viewed as a generalization of the following criteria:
von Mises, Drucker-Prager, Tresca, modified Tresca, Coulomb-Mohr, modified Cam-clay, Deshpande and Fleck (2000),
Rankine, and Ottosen (1977) (the last two 
for the deviatoric section). 
Obviously, the criterion may account for situations which cannot be described by the simple criteria to which it
reduces in particular cases. Several examples of this may be found in the field of granular media, where
several {\it ad hoc} yield conditions have been proposed, which may describe {\it one peculiar
material, but cannot describe another}. In the present paper, it is shown with several examples 
that our yield criterion provides a 
unified description for a extremely broad class of quasi-brittle and frictional materials.
Beyond the evident interest in generalization, there is a specific motivation for advocating 
the necessity of having a single criterion describing different materials. This lies in the fact that {\it during 
hardening, a yield surface may evolve from the shape typical of a certain material to that typical of another}.
An evident example of this behaviour can be found in the field of granular materials, referring in particular to 
metal powders. These powders become cohesive during compaction, so that the material is initially a true granular 
material, but becomes finally a porous metal, whose porosity may be almost completely eliminated through sintering.
The key to simulate this process is plasticity theory, so that a yield function must be employed  
evolving from the typical shape of a granular material (`triangular' deviatoric and 
`drop-shaped' meridian sections), to that of a porous metal (circular deviatoric and elliptic meridian sections)
and, in case of sintering, to that of a fully-dense metal (von Mises criterion).
Another example of extreme shape variation of yield function during hardening is the process of decohesion of a
rock-like material due to damage accumulation, a situation in a sense opposite to that described above.
Evidently, a continuous distortion of the yield surface can be described employing the criterion proposed
in this paper and simply making material parameters depend on hardening.

\section{A premise on Haigh-Westergaard representation}

\begin{figure}[!htcb]
  \begin{center}
      \includegraphics[width= 8 cm]{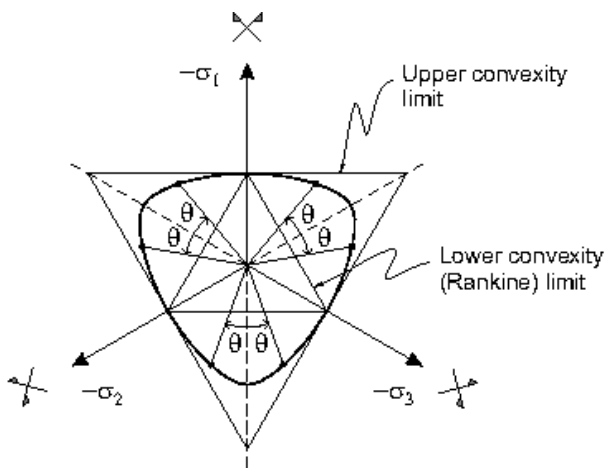}
\caption{\footnotesize Deviatoric section: definition of angle $\theta$, symmetries, 
lower and upper convexity bounds.}
\label{fig1}
  \end{center}
\end{figure}

The analysis will be restricted to isotropic behaviour, therefore the Haigh-Westergaard
representation of the yield locus is employed (Hill, 1950). This
is well-known, so that we limit the presentation here to a few remarks that may be 
useful in the following. First, we recall that:
\begin{description}
\item[${\mathcal A}1.$] a single point in the Haigh-Westergaard space is representative of the infinite
(to the power three) stress tensors having the same principal values;
\item[${\mathcal A}2.$] due to the arbitrariness in the ordering of the eigenvalues of a tensor, 
six different points correspond in the Haigh-Westergaard representation to a given stress tensor.
As a result, the yield surface results symmetric about the projections of the principal axes 
on the deviatoric plane (Fig. \ref{fig1});
\item[${\mathcal A}3.$] the Haigh-Westergaard representation preserves the scalar product only
between coaxial tensors;
\item[${\mathcal A}4.$] a convex yield surface ---for a material with a fixed yield 
strength under triaxial compression--- must be internal to the two limit situations shown in 
Fig.~\ref{fig1} (Haythornthwaite, 1985). Note 
that the inner bound will be referred as {\lq}the Rankine limit{\rq}.
\end{description}
Due to isotropy, the analysis of yielding can be pursued fixing once and for all 
a reference system 
and restricting to all stress tensors diagonal in this system. We will refer to this setting
as to the Haigh-Westergaard representation. When 
tensors (for instance,
the yield function gradient) coaxial to the reference system are represented, 
the scalar product is preserved, property ${\mathcal A}3$.
In the Haigh-Westergaard representation, the hydrostatic and deviatoric stress components are 
defined by the invariants 
\beq\lb{eq: fi}
p = -\frac{\tr \stress}{3},~~~q = \sqrt{3 J_2},
\eeq
where 
\beq\lb{eq: ga}
J_2 = \frac{1}{2} \devstress\scalp\devstress,
~~~\devstress = \stress - \frac{\tr \stress}{3}\Id,
\eeq
in which $\devstress$ is the deviatoric stress, $\Id$ is the identity tensor, a dot denotes 
scalar product and
$\tr$ denotes the trace operator, so that $\bA\scalp\bB = \tr \bA\bB^T$, for every second-order
tensors $\bA$ and $\bB$. The position
of the stress point in the deviatoric plane is singled out by the Lode (1926) angle $\theta$ defined as
\beq
\lb{invarianti2}
\theta = \frac{1}{3} \, \cos^{-1} \left(
\frac{3\sqrt{3}}{2}
\frac{\Jt}{\Js^{3/2}} \right),~~~
J_3 = \frac{1}{3} \, \tr \devstress^3,
\eeq
so that $\theta \in [0, \pi/3]$. As a consequence of property $({\mathcal A}2)$ 
of the Haigh-Westergaard representation, a single value of $\theta$ corresponds to
six different points in the deviatoric plane (Fig. \ref{fig1}).
The following gradients of the invariants, that will be useful later, 
\beq
\lb{derivateinv}
\begin{array}{l}
\ds{\frac{\partial p}{\partial \stress} = -\frac{1}{3} \Id,~~\frac{\partial J_2}{\partial \stress} = \bS,~~
\frac{\partial J_3}{\partial \stress} = \bS^2 - \frac{\tr \bS^2}{3}\Id,} \\ [5 mm]
\ds{\frac{\partial \theta}{\partial \stress} = -\frac{9}{2\,q^3 \, \sin 3\theta} \left(
\bS^2 -\frac{\tr \bS^2}{3} \Id - q \frac{\cos 3\theta}{3} \bS
\right)},
\end{array}
\eeq
can be obtained from well-known formulae (e.g. Truesdell and Noll, 1965, Sect. 9) using the identity
\beq
\frac{\partial \bS}{\partial \stress} = \Id \bob \Id - \frac{1}{3} \Id \otimes \Id,
\eeq
where the symbol $\otimes$ denotes the usual dyadic product and 
$\Id\bob\Id$ is the symmetrizing fourth-order tensor, defined for every tensor $\bA$ as
$\Id\bob\Id[\bA] = (\bA+\bA^T)/2$. Note that $\partial \theta/\partial \stress$ is orthogonal to $\Id$
and to the deviatoric stress $\bS$. 

\section{A new yield function}

We propose the seven-parameters yield function 
$\yf:\,\Sym\rightarrow\mathbb{R}\cup\{+\infty\}$ defined as:
\beq
\lb{eq: yield function}
\yf(\stress)= f(p) +\frac{q}{g(\theta)},
\eeq
where the dependence on the stress $\stress$ is included in the invariants $p$, $q$ and
$\theta$, eqns (\ref{eq: fi}) and (\ref{invarianti2}), through the `meridian' function
\beq\lb{effedip}
f(p) =\left\{
\barr{ll}
-Mp_c\sqrt{\left(\Phi-\Phi^m\right)
\left[2(1-\alpha)\Phi+\alpha\right] }
&~~~\mathit{if}\ \Phi\in[0,1],\\ [3 mm]
+\infty&~~~\mathit{if}\ \Phi\notin[0,1],
\earr\right.
\eeq
where
\beq\lb{fiegi}
\Phi=\frac{p+c}{p_c+c},
\eeq
describing the pressure-sensitivity\footnote{The meridian function can be written in an alternative form 
by using the Macauley bracket operator, defined for every scalar $\alpha$ as $<\alpha> = \max\{0, \alpha\}$,
and the indicator function $\chi_{[0,1]}(\Phi)$, which takes the value 0 when 
$\Phi \in [0,1]$ and is equal to $+\infty$ otherwise
$$
f(p) =
-Mp_c\sqrt{\left(\tilde\Phi-\tilde\Phi^m\right)
\left[2(1-\alpha)\tilde\Phi+\alpha\right]} + \chi_{[0,1]}(\Phi),~~~\tilde\Phi = <\Phi>-<\Phi -1>.
$$}
and the `deviatoric' function
\beq\lb{gi}
g(\theta)= \frac{1}{\cos{\left[ \beta \frac{\pi}{6} - \frac{1}{3}\cos^{-1}\left(\gamma \cos{3 \theta}\right)\right]}},
\eeq
describing the Lode-dependence of yielding.
The seven, non-negative material parameters: 
\beq
\lb{sevenstones}
\underbrace{M > 0,~ p_c > 0,~ c \geq 0,~ 0<\alpha < 2,~ m > 1}_{\mbox{defining}~\ds{f(p)}},~~~ 
\underbrace{0\leq \beta \leq 2,~ 0 \leq \gamma < 1}_{\mbox{defining}~\ds{g(\theta)}}, 
\eeq
define the shape
of the associated (single, smooth) yield surface. 
In particular, $M$ controls the pressure-sensitivity, $p_c$ and $c$ are the yield strengths under isotropic 
compression and tension, respectively. Parameters $\alpha$ and $m$ define the distortion of the meridian section, 
whereas $\beta$ and $\gamma$ model the shape of the deviatoric section. Note that the deviatoric function describes
a piecewise linear deviatoric surface in the limit $\gamma \longrightarrow 1$.
Finally, it is important to remark that within the interval of $\beta \in [0, 2]$ the yield function
is convex independently of the values assumed by parameter $\gamma$.
Convexity requirements, that will be proved later, impose a broader variation of $\beta$ than (\ref{sevenstones})$_6$, 
but the interval where $\beta$ may range becomes a function of $\gamma$.
In particular, the yield function is convex when
\beq
\lb{sevenplus}
2-\mathcal{B}(\gamma) \leq \beta \leq \mathcal{B}(\gamma),
\eeq
where function $\mathcal{B}(\gamma)$ takes values within the interval $]2, 4]$, when $\gamma$ ranges in $[0, 1[$ 
and is defined as
\beq
\mathcal{B}(\gamma) = \left. 3-\frac{6}{\pi} \tan^{-1} \frac{1-2 \cos z -2 \cos^2 z}{2 \sin z (1-\cos z)
}\right|_{z=2/3(\pi-\cos^{-1}\gamma)}.
\eeq

The yield function (\ref{eq: yield function}) corresponds to the following yield surface:
\beq\lb{eq: yield surface 2}
q=-f(p)g(\theta),~~~
~p\in[-c,p_c],~~\HWt\in[0,\pi/3],
\eeq
which makes explicit the fact that $f(p)$ and $g(\theta)$ define the shape of the
meridian and deviatoric sections, respectively.

\begin{figure}[!htcb]
  \begin{center}
      \includegraphics[width= 16 cm]{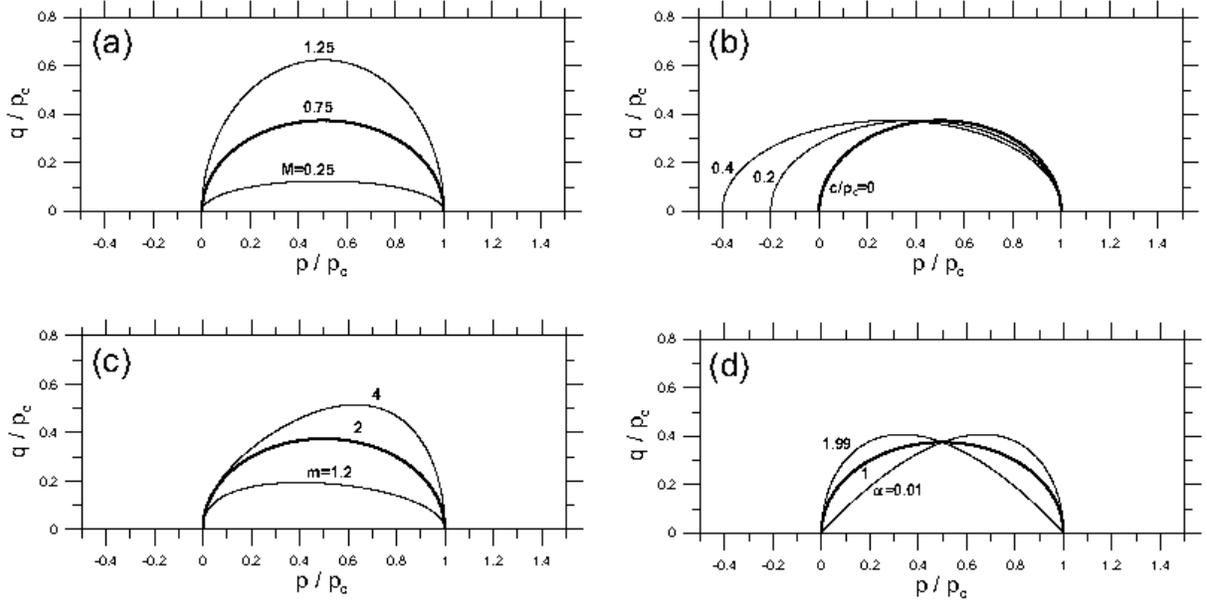}
\caption{\footnotesize Meridian section: effects related to the variation of parameters 
$M$ (a), $c/p_c$ (b),
$m$ (c), and $\alpha$ (d).}
\label{fig2}
  \end{center}
\end{figure}

\begin{figure}[!htcb]
  \begin{center}
      \includegraphics[width= 12 cm]{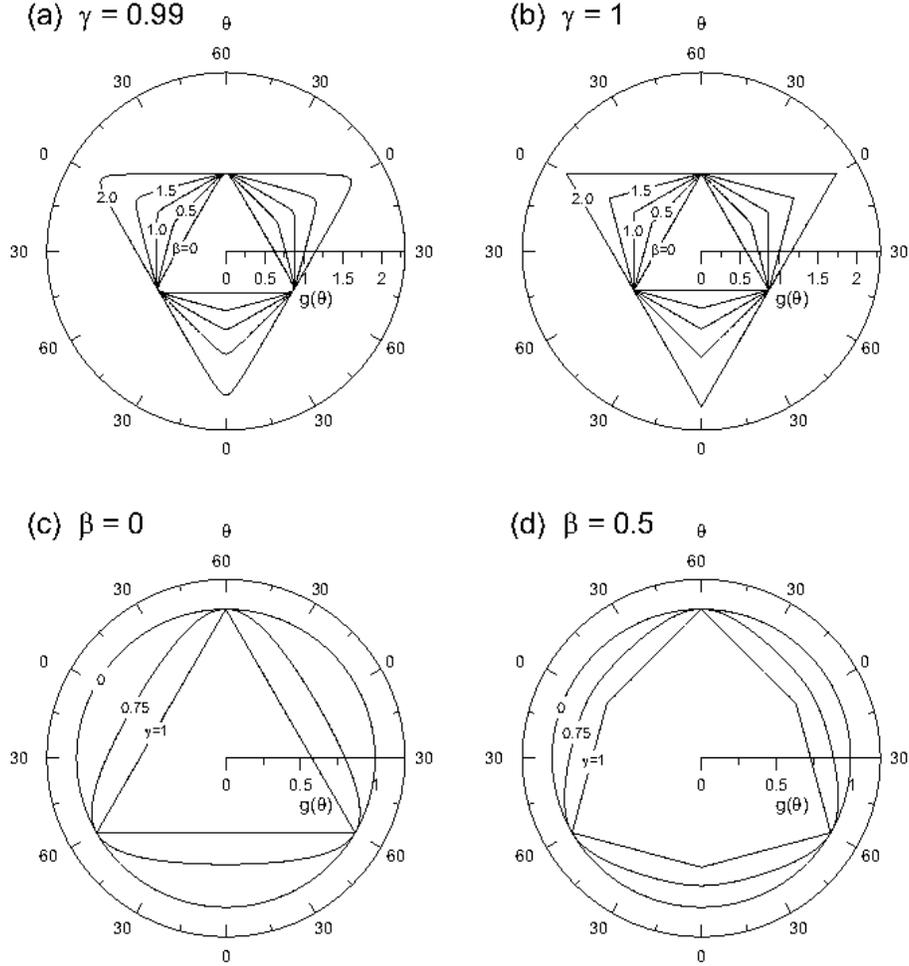}
\caption{\footnotesize Deviatoric section: 
effects related to the variation of $\beta$ and $\gamma$. 
Variation of $\beta = 0, 0.5, 1, 1.5, 2$ at fixed $\gamma= 0.99$ (a) and $\gamma=1$ (b).
Variation of $\gamma = 1, 0.75, 0$ at fixed 
$\beta = 0$ (c) and $\beta=0.5$ (d).}
\label{fig3}
  \end{center}
\end{figure}

The yield surface (\ref{eq: yield surface 2}) is sketched in Figs.~\ref{fig2}-\ref{fig3}
for different values of the seven above-defined material parameters (non-dimensionalization is
introduced through division by $p_c$ in Fig.~\ref{fig2}). 
In particular, meridian sections are reported in Figs.
\ref{fig2} ($g(\theta)=1$ has been taken), whereas Fig. \ref{fig3} pertains to deviatoric sections.
 
As a reference, the case corresponding to the 
modified Cam-clay introduced by Roscoe and Burland (1968) and Schofield and Wroth (1968) and corresponding to 
$\beta=1$, $\gamma=0$, $\alpha=1$, $m=2$, and $c=0$  
is reported in Fig. \ref{fig2} as a solid line, for $M = 0.75$. 
The distortion of meridian section reported in Fig. 
\ref{fig2} (a) ---where $M = 0.25, 0.75, 1.25$--- can also be obtained within the framework of the
modified Cam-clay, 
whereas the effect of
an increase in cohesion reported in Fig. \ref{fig2} (b) ---where $c/p_c = 0, 0.2, 0.4$--- 
may be employed
to model the gain in cohesion consequent to plastic strain, during compaction of powders.

The shape distortion induced by the variation of parameters 
$m$ and $\alpha$, Fig. \ref{fig2} (c) ---where $m = 1.2, 2, 4$--- and (d) 
---where $\alpha = 0.01, 1.00, 1.99$--- 
is crucial to fit experimental results relative to
frictional materials. 

A unique feature of the proposed model is 
the possibility of extreme shape distortion of the deviatoric section, which may range between the 
upper and lower convexity limits, and approach Tresca, von Mises and Coulomb-Mohr. This 
is sketched in Fig.~\ref{fig3}, where to simplify reading of the figure,
function $g(\theta)$ has been normalized through division by $g(\pi/3)$, so that all deviatoric
sections coincide at the point $\theta = \pi/3$. 
The use of our model may therefore allow one to simply obtain a convex, smooth approximation of
several yielding criteria (Tresca and Coulomb-Mohr, for instance). 
If this may be not substantial from theoretical point of view, it clearly avoids 
the necessity of introducing independent yielding mechanisms.

Parameter $\gamma$ is kept fixed in Figs.~\ref{fig3} (a) and (b) and equal to 0.99 and 1, respectively,
whereas parameter $\beta$ is fixed in Figs.~\ref{fig3} (c) and (d) and equal to 0 and 1/2.
Therefore, figures (a) and (b) demonstrate the effect of the variation in $\beta$ (= 0, 0.5, 1, 1.5, 2)
which makes possible a distortion of the yield surface from the upper to lower convexity limits
going through Tresca and Coulomb-Mohr shapes. The role played by $\gamma$ (= 1, 0.75, 0) 
is investigated in figures 
(c) and (d), from which it becomes evident that $\gamma$ has a smoothing effect on the 
corners, emerging in the limit $\gamma=1$. The von Mises (circular) deviatoric section emerges
when $\gamma=0$.

\begin{figure}[!htcb]
  \begin{center}
      \includegraphics[width= 12 cm]{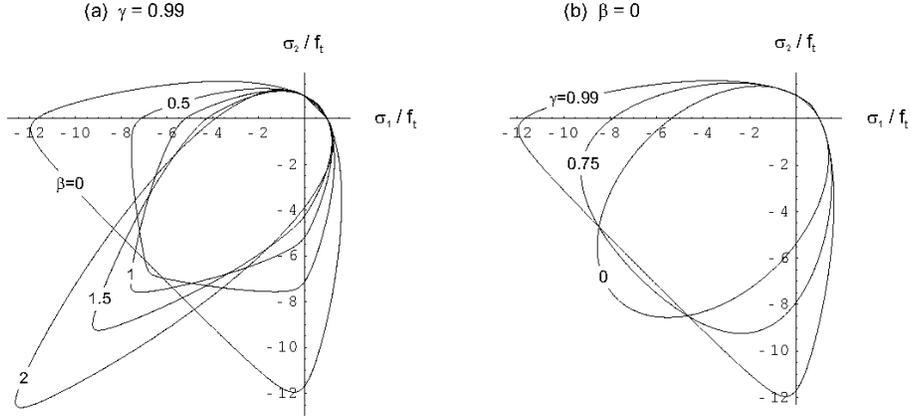}
\caption{\footnotesize Yield surface in the biaxial plane $\sigma_1/f_t$ vs. $\sigma_2/f_t$, with $\sigma_3=0$.
Variation of $\beta = 0, 0.5, 1, 1.5, 2$ at fixed $\gamma= 0.99$ (a) and 
variation of $\gamma = 0, 0.75, 0.99$ at fixed $\beta = 0$ (b).}
\label{fig4}
  \end{center}
\end{figure}

The yield surface in the biaxial plane 
$\sigma_1$ versus $\sigma_2$, with $\sigma_3=0$ is sketched in Fig.~\ref{fig4}, where axes are 
normalized through division by the uniaxial tensile strength $f_t$. In particular, the figure pertains to 
$M=0.75$, $p_c = 50 c$, $m=2$, and $\alpha = 1$, whereas $\gamma = 0.99$ 
is fixed and $\beta$ is equal to \{0, 0.5, 1, 1.5, 2\} in 
Fig.~\ref{fig4} (a) and, vice-versa, $\beta = 0$ is fixed and $\gamma$ is equal to \{0, 0.75, 0.99\} in 
Fig.~\ref{fig4} (b).

\subsection{Smoothness of the yield surface}

Smoothness of yield surface (\ref{eq: yield surface 2}) within the interval of material parameters defined in 
(\ref{sevenstones})-(\ref{sevenplus}) can be proved considering the 
yield function gradient. This can be obtained from (\ref{derivateinv})
in the form
\beq
\lb{grad}
\frac{\partial F}{\partial \stress} = 
a(p)\, \Id + b(\theta)\, \tilde{\bS} + c(\theta)\, \tilde{\bS}^{\perp},
\eeq
where
\beq
\tilde{\bS} = \sqrt{\frac{3}{2}}\frac{\bS}{q},~~~~
\tilde{\bS}^{\perp} = -\frac{\sqrt{2}}{\sqrt{3}\, q} \frac{\partial \theta}{\partial \stress} = 
\frac{1}{\sin 3 \theta}\left[\sqrt{6}\left(\tilde{\bS}^{2}-\frac{1}{3} \Id\right)
-\cos 3\theta\, \tilde{\bS} \right],
\eeq
and
\beqar
a(p) &=& -\frac{1}{3}\derivative{f(p)}{p} = \frac{Mp_c}{3(p_c+c)}
                                  \frac{\left(1-m\Phi^{m-1}\right)
                                  \left[2(1-\alpha)\Phi+\alpha\right]+
                                  2(1-\alpha)\left(\Phi-\Phi^m\right)}
                                  {2\sqrt{\left(\Phi-\Phi^m\right)
                                  \left[2(1-\alpha)\Phi+\alpha\right]}}, \nonumber \\
b(\theta) &=& \sqrt{\frac{3}{2}} \frac{1}{g(\theta)}, \\
c(\theta) &=& -\frac{\sqrt{3}\gamma\sin 3 \theta}{\sqrt{2}\sqrt{1-\gamma^{2}\cos^{2} 3\theta}}\sin\left[\beta\frac{\pi}{6}-\frac{1}{3}\cos^{-1}(\gamma\cos 3\theta)\right]. \nonumber
\eeqar
It should be noted that $c(0)=c(\pi/3)=0$ and that 
$\tilde{\bS}$ and $\tilde{\bS}^\perp$ are unit norm, coaxial and normal to each other tensors\footnote{
Note that $c \, \tilde{\bS}^{\perp} = \b0$ at $\theta = 0, \pi/3$. 
This can be deduced from the fact that $|\tilde{\bS}^{\perp}| = 1$ and $c=0$ for 
$\theta = 0, \pi/3$ or, alternatively, can be proved directly observing that for 
$\theta = 0, \pi/3$ the deviatoric stress can be generically written as 
$\{S_1, -S_1/2, -S_1/2\}$, unless all (uninfluent) permutations of components.
}. 
Coaxiality and orthogonality are immediate properties, whereas the proof that $|\tilde{\bS}^\perp| = 1$ 
is facilitated when 
the following identities are kept into account
\beq
\tilde{\bS}^3-\frac{1}{2}\tilde{\bS}- \frac{\cos 3 \theta}{3\sqrt{6}}\Id = \b0, ~~\leadsto ~~
\tilde{\bS}^2\scalp\tilde{\bS}^2 = \frac{1}{2},
\eeq
the former of which is the Cayley-Hamilton theorem written for $\tilde{\bS}$.
Let us consider now from (\ref{grad}) the unit-norm yield function gradient 
\beq
\lb{unitq}
\gradF =
\frac{a}{\sqrt{3a^2+b^2+c^2}}\, \Id + \frac{b}{\sqrt{3a^2+b^2+c^2}}\, \tilde{\bS} + 
\frac{c}{\sqrt{3a^2+b^2+c^2}}\, \tilde{\bS}^{\perp},
\eeq
defining, for stress states satisfying $F(\stress)=0$, the unit normal to the yield surface.
The following limits can be easily calculated 
\beq
\lb{diciassette}
\lim_{\Phi \rightarrow 0^+} \gradF = \frac{1}{\sqrt{3}}\Id,~~~
\lim_{\Phi \rightarrow 1^-} \gradF = -\frac{1}{\sqrt{3}}\Id,
\eeq
so that the yield surface results to be smooth at the limit points where the hydrostatic axis is met. Moreover, 
smoothness of the deviatoric section of the yield surface is proved observing that 
\beq
\lb{devlim}
\lim_{\theta \rightarrow 0, \pi/3} 
\gradF = \frac{a}{\sqrt{3a^2+b^2}}\, \Id + \frac{b}{\sqrt{3a^2+b^2}}\, \tilde{\bS},
\eeq
where $\tilde{\bS}$ and $b$ are evaluated at $\theta = 0$ and $\theta = \pi/3$, and noting that 
$\tilde{\devstress}$ and $\tilde{\devstress}^\perp$ are coaxial, deviatoric tensors so that they are represented by  two orthogonal vectors
in the deviatoric plane in the Haigh-Westergaard stress space.
We observe, finally, that limits (\ref{diciassette}) do not hold true when $\alpha$ equals 0 and 2 and that 
limits (\ref{devlim}) does not hold true when $\gamma = 1$. In particular, 
a corner appears at the 
intersection of the yield surface with the hydrostatic axis in the former case and the 
deviatoric section
becomes piecewise linear in the latter.

\subsection{Reduction of yield criterion to known cases}

The yield function (\ref{eq: yield function})-(\ref{gi}) reduces to almost all\footnote{A remarkable exception is the 
isotropic Hill (1950 b) criterion, corresponding to a Tresca criterion rotated 
of $\pi/6$ in the deviatoric plane.} 
`classical' criteria of yielding. These can be obtained as limit cases in the way illustrated in Tab.~\ref{limiti} 
where the modified Tresca criterion was introduced by Drucker (1953), 
whereas the Haigh-Westergaard representation of the Coulomb-Mohr criterion was proposed by Shield (1955).
In Tab.~\ref{limiti} parameter $r$ denotes the ratio between the uniaxial strengths in compression (taken positive) and tension, indicated
by $f_c$ and $f_t$, respectively. We note that for real materials $r \geq 1$ and that we did not
explicitly consider the special cases of no-tension $f_t=0$ or granular $f_t=f_c=0$ materials [which anyway can be
easily incorporated as limits of (\ref{eq: yield function})-(\ref{gi})].

We note that the expression of the Tresca criterion which follows from (\ref{eq: yield function})-(\ref{gi}) in the 
limits specified in Tab.~\ref{limiti}, was provided also by Bardet (1990) and 
answers ---in a positive way--- 
the question (raised by Salen\c{c}on, 1974) if a proper\footnote{The expression 
$$
f(\stress) = 4 J_2^3 - 27 J_3^2 - 36 k^2J_2^2 + 96 k^4 J_2 - 64 k^6,
$$
where $k$ is the yield stress under shear (i.e. $k=f_t/2$), reported in several textbooks on plasticity,
is definitively wrong. This can be easily verified taking a stress state belonging to one of the planes defining the
Tresca criterion, but outside the yield locus, for instance, the point $\{\sigma_1= 0,~\sigma_2 = -2k,~\sigma_3 = 2k\}$,
corresponding to $J_2=4 k^2$ and $J_3 = 0$. Obviously, the point lies well outside the yield locus, but satisfies
$f(\stress)=0$, when the above, wrong, yield function is used.
}
form of the criterion in terms of stress invariants exist.

The Mohr-Coulomb limit merits a special mention. In fact, if the following values of the parameters are selected
\beq
\alpha=0,~~~
\ds{ c=\frac{f_c \left[ \cos\left(\beta\frac{\pi}{6}-\frac{\pi}{3}\right)+\cos \beta\frac{\pi}{6}\right]}{3r\cos\left(\beta\frac{\pi}{6}-\frac{\pi}{3}\right)
-3\cos\beta\frac{\pi}{6}} }, ~~~
\ds{M=\frac{3\left[r\cos\left(\beta\frac{\pi}{6}-\frac{\pi}{3}\right)-\cos\beta\frac{\pi}{6}\right]}{\sqrt{2}(r+1)}},
\eeq
and then the limits 
\beq
\gamma \longrightarrow 1,~~~p_c=f_c\,m \longrightarrow \infty,
\eeq
are performed, a three-parameters generalization of Coulomb-Mohr criterion is obtained, 
which reduces to the latter criterion
in the special case when $\beta$ is selected in the form specified in Tab.~\ref{limiti} (yielding an expression 
noted also by Chen and Saleeb, 1982).
\begin{table}[!htcb]
\begin{center}
\caption{\footnotesize Yield criteria obtained as special cases of (\ref{eq: yield function})-(\ref{gi}),
$r=f_c/f_t$ and $f_c$ and $f_t$ are the uniaxial strengths in compression and tension, respectively.}
\vspace{3 mm}
\label{limiti}
\begin{tabular}{||l|c|c||}
\hline\hline 
 & &\\ [-2 mm]
Criterion   & Meridian function $f(p)$ & Deviatoric function $g(\theta)$            \\  [3 mm]
\hline\hline
von Mises            & $\begin{array}{ll} ~&~ \\ \ds{\alpha=1,} & \ds{m = 2,} \\ \ds{M=\frac{2 f_t}{p_c}}, & c=p_c= \longrightarrow \infty \\ ~&~ \end{array} $& $\beta=1$, $\gamma=0$                  \\
\hline
Drucker-Prager       & $\begin{array}{ll} ~&~ \\ \ds{\alpha=0}, & \ds{M=\frac{3(r-1)}{\sqrt{2}(r+1)}}, \\ \ds{c=\frac{2 f_c}{3(r-1)}}, & p_c=f_c\,m \longrightarrow \infty \\ ~&~ \end{array}$  & as for von Mises        \\
\hline
Tresca               &$\begin{array}{c} ~ \\ \mbox{as for von Mises, except that} \\ \ds{M=\frac{\sqrt{3}f_t}{p_c}} \\ ~  \end{array}$ 
& $\beta=1$, $\gamma \longrightarrow 1$                  \\
\hline
mod. Tresca          & $\begin{array}{c} ~ \\ \mbox{as for Drucker-Prager, except that} \\ [-3 mm] ~ \\ \ds{M=\frac{3\sqrt{3}(r-1)}{2\sqrt{2}(r+1)}} \\ ~  \end{array}$ & as for Tresca                   \\
\hline
Coulomb-Mohr         & $\begin{array}{c} ~ \\ \mbox{as for Drucker-Prager, except that}  \\ \ds{M=\frac{3\left[r\cos\left(\beta\frac{\pi}{6}-\frac{\pi}{3}\right)-\cos\beta\frac{\pi}{6}\right]}{\sqrt{2}(r+1)}}
\\ [-3 mm] ~ \\
\ds{ c=\frac{f_c \left[\cos\left(\beta\frac{\pi}{6}-\frac{\pi}{3}\right) + \cos \beta\frac{\pi}{6}\right]}{3r\cos\left(\beta\frac{\pi}{6}-\frac{\pi}{3}\right)
-3\cos\beta\frac{\pi}{6}} }\\ ~  \end{array}$ 
& $\begin{array}{c}  \ds{\beta=\frac{6}{\pi}\tan^{-1}\frac{\sqrt{3}}{2r+1}},\\~ \\ \gamma \longrightarrow 1 \end{array} $ \\
\hline
mod. Cam-clay        & $\begin{array}{c} ~ \\ m=2,~\alpha=1,~c=0 \\ ~ \end{array}$ & as for von Mises           \\
\hline\hline
\end{tabular}
\end{center}
\end{table}
The cases reported in Tab.~\ref{limiti} refer to situations in which the 
criterion (\ref{eq: yield function})-(\ref{gi}) reduces to known yield criteria both in terms of function
$f(p)$ and of function $g(\theta)$. 
It is however important to mention that the Lode's dependence function $g(\theta)$ 
reduces also to well-known cases, but in which the pressure-sensitivity cannot be described by the meridian function 
(\ref{effedip}). These are reported in Tab.~\ref{limitidev}. It is important to mention that the form of
our function $g(\theta)$, eqn. (\ref{gi}), was indeed constructed as a generalization of
the deviatoric function introduced by Ottosen (1977).
\begin{table}[!htcb]
\begin{center}
\caption{\footnotesize Deviatoric yield functions obtained as special cases of (\ref{gi})}
\vspace{3 mm}
\label{limitidev}
\begin{tabular}{||l|l||}
\hline\hline 
Criterion  & Deviatoric function $g(\theta)$    \\
\hline\hline
Lower convexity (Rankine)& $\beta=0$, ~$\gamma \longrightarrow 1$           \\
\hline
Upper convexity      & $\beta=2$, ~$\gamma \longrightarrow 1$           \\
\hline
Ottosen              & $\beta=0$, ~$0\leq \gamma < 1$ \\
\hline\hline
\end{tabular}
\end{center}
\end{table}

\subsection{A comparison with experiments}
\lb{fitting}
A brief comparison with experimental results referred to several materials is reported below. 
We limit the presentation to a few representative examples demonstrating the extreme flexibility of 
the proposed model to fit experimental results. 
In particular, we concentrate on the meridian section, 
whereas only few examples are provided for the deviatoric section, which has a shape so deformable and
ranging between well-known forms that fitting experiments is a-priori expected. Results on the
biaxial plane $\sigma_1-\sigma_2$ are also included. All values of material parameters defining 
the yield function (\ref{eq: yield function})-(\ref{gi}) employed to fit experimental data 
may be useful as a reference and are reported in Appendix A.

\begin{figure}[!htcb]
  \begin{center}
      \includegraphics[width= 12 cm]{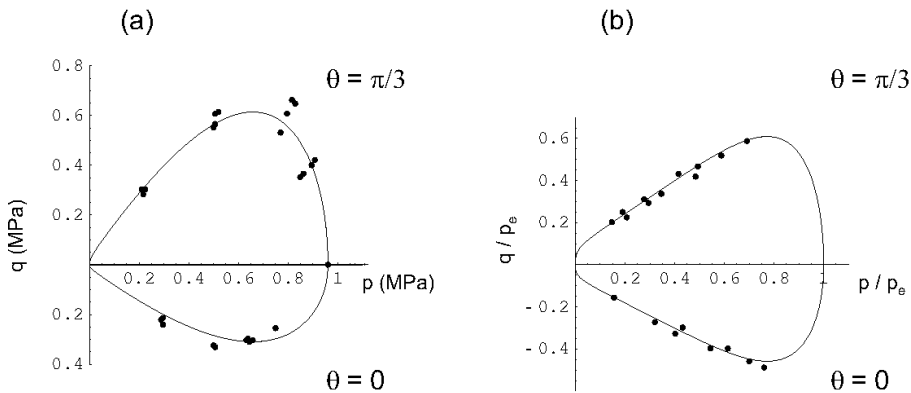}
\caption{\footnotesize Comparison with experimental results relative to sand (Yasufuku et al. 1991) (a)
and clay (Parry, reported by Wood, 1990) (b).}
\label{fig5}
  \end{center}
\end{figure}

Typical of soils are the experimental results reported in Fig. \ref{fig5}, on Aio dry sand and 
Weald clay, taken, respectively, from Yasufuku et al. (1991, their Fig. 10a) and Parry 
(reported by Wood, 1990, their Fig. 7.22, so that $p_e$ is the equivalent consolidation 
pressure in Fig.~\ref{fig5}(b)). 
Note that the upper plane of
the graphs refers to triaxial compression ($\theta= \pi/3$), whereas triaxial extension is reported in
the lower part of the graphs ($\theta= 0$).
It may be concluded from the figure that
experimental results can be easily fitted by our function $f(p)$, still maintaining 
a smooth intersection of the yield surface with $p-$axis.

\begin{figure}[!htcb]
  \begin{center}
      \includegraphics[width= 12 cm]{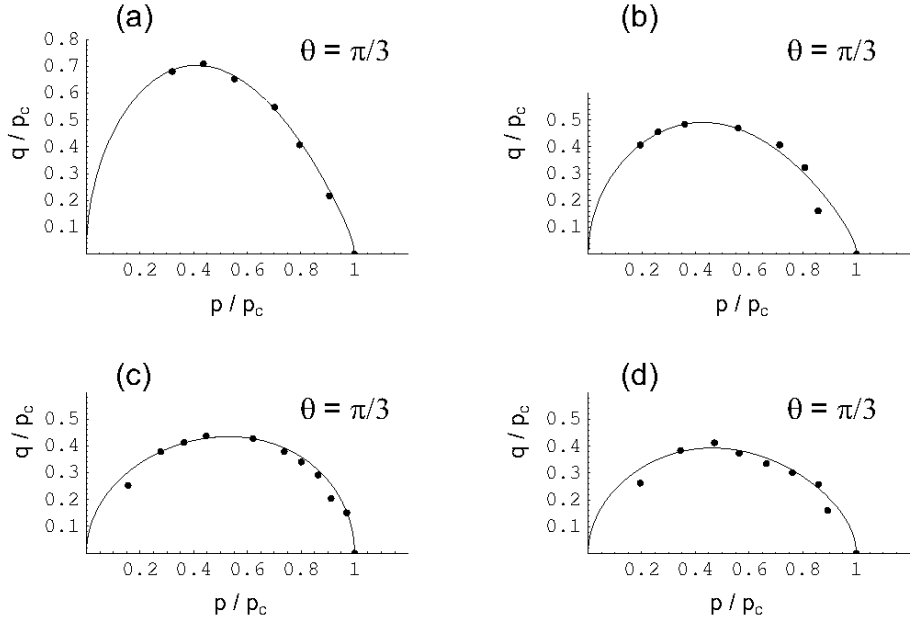}
\caption{\footnotesize Comparison with experimental results relative to aluminium
powder (a) aluminum composite powder (b), lead powder (c) and lead shot-steel composite 
powder(d), 
data taken from Sridhar and Fleck (2000).}
\label{fig6}
  \end{center}
\end{figure}

In addition to soils, the proposed function (\ref{eq: yield function})-(\ref{gi}) can model yielding of porous 
ductile or cellular materials,
metallic and composite powders, concrete and rocks.
To further develop this point, a comparison with experimental results given by Sridhar and Fleck (2000) 
---their Figs. 5(b) and 9(c)--- relative 
to ductile powders is reported in Fig.~\ref{fig6}. In particular, Fig.~\ref{fig6} (a) is relative to an aluminum powder 
(Al D$_0$=0.67, D=0.81 in Sridhar and Fleck, their Fig. 5b), Fig.~\ref{fig6} 
(b) to an aluminum powder reinforced by 40 vol.\%SiC (Al-40\%Sic D$_0$=0.66, D=0.82 in Sridhar and Fleck, their Fig. 5b), 
Fig.~\ref{fig6} (c) to a lead powder (0\% steel in Sridhar and Fleck, their Fig. 9c),
and Fig.~\ref{fig6} (d) to a lead shot-steel composite powder (20\% steel in Sridhar and Fleck, their Fig. 9c).
Beside the fairly good agreement between experiments and proposed yield function, we note
that the aluminium powder has a behaviour ---different from soils and lead-based powders--- resulting in
a meridian section of the yield surface similar to the early version of the Cam-clay model (Roscoe and
Schofield, 1963).

\begin{figure}[!htcb]
  \begin{center}
      \includegraphics[width= 12 cm]{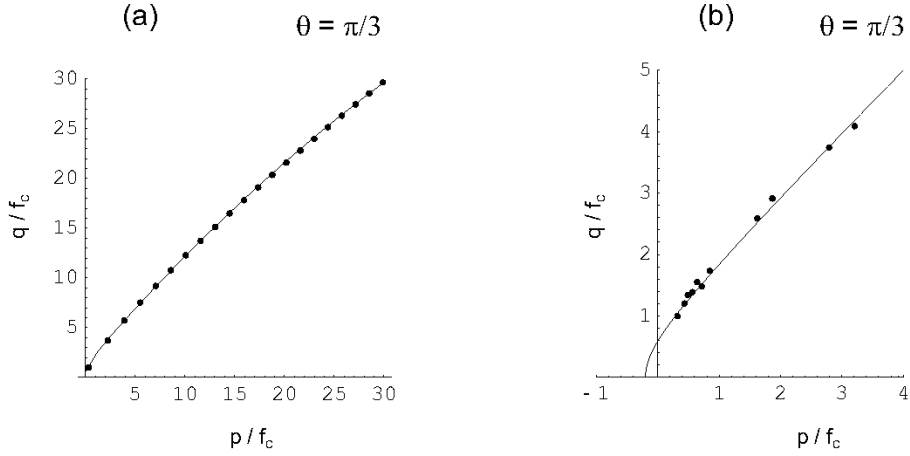}
\caption{\footnotesize Comparison with the experimental relation (\ref{newman}) proposed
by Newman and Newman (1971) (a) and with results by Sfer et al. (2002) (b).}
\label{fig7}
  \end{center}
\end{figure}

Regarding concrete, among the many experimental results currently available, 
we have referred to Sfer et al.  (2002, their Fig. 6) and to the Newman and Newman (1971) empirical relationship
\beq
\lb{newman}
\frac{\sigma_1}{f_c} = 1 + 3.7 \left( \frac{\sigma_3}{f_c} \right)^{0.86},
\eeq
where $\sigma_1$ and $\sigma_3$ are the maximum and minimum principal stresses at failure and
$f_c$ is the value of the ultimate uniaxial compressive strength.
Small circles in Fig.~\ref{fig7} represents results obtained using relationship (\ref{newman}) in figure (a) and
experimental results by Sfer et al. (2002) in figure (b);
the approximation provided by the criterion (\ref{effedip})-(\ref{fiegi}) is also reported as a continuous line.

\begin{figure}[!htcb]
  \begin{center}
      \includegraphics[width= 9 cm]{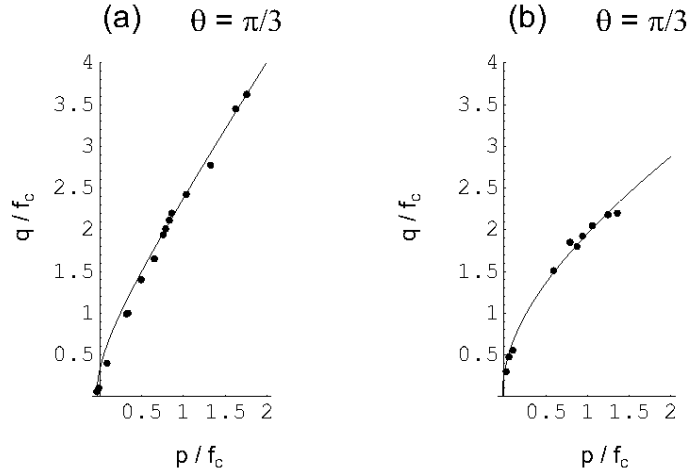}
\caption{\footnotesize Comparison with experiments for rocks. Chert (a) and dolomite (b), 
data taken from Hoek and Brown (1980).}
\label{fig8}
  \end{center}
\end{figure}

As far as rocks are concerned, we limit to a few examples. However, we believe that due to the fact that our
criterion approaches Coulomb-Mohr, it should be particularly suited for these materials. 
In particular,
data taken from Hoek and Brown (1980, their pages. 143 and 144) are reported in Fig.~\ref{fig8} as 
small circles for two rocks, chert (Fig.~\ref{fig8} a) and dolomite (Fig.~\ref{fig8} b).

\begin{figure}[!htcb]
  \begin{center}
      \includegraphics[width= 12 cm]{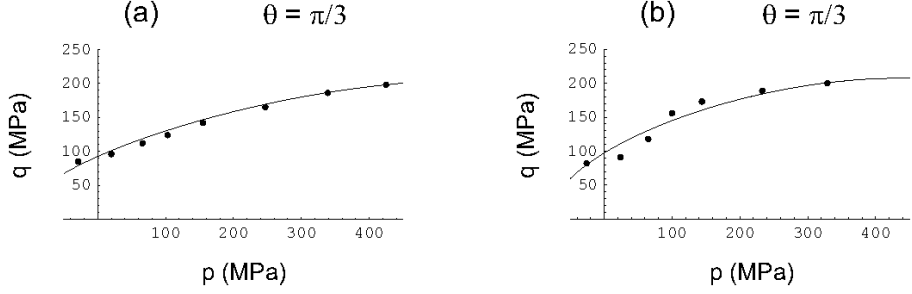}
\caption{\footnotesize Comparison with experimental results for polymers. Methacrylate (a)
and an epoxy binder (b), data taken from Ol'khovik (1983).}
\label{fig9}
  \end{center}
\end{figure}

A few data on polymers are reported in Fig.~\ref{fig9} ---together with the fitting provided by our model--- 
concerning polymethil methacrylate (Fig.~\ref{fig9} a)
and an epoxy binder (Fig.~\ref{fig9} b), taken from Ol'khovik (1983, their Fig. 5), see also 
Altenbach and Tushtev (2001, their Figs. 2 and 3).
 
Finally, our model describes ---with a different yield function--- the same yield surface proposed 
by Deshpande and Fleck (2000) to describe the behaviour of metallic foams. 
In particular, the correspondence between parameters of our model (\ref{eq: yield function})-(\ref{gi})
and of the yield surface proposed by Deshpande and Fleck [2000, their eqns. (2)-(3)] is obtained setting
$$
\beta=1,~~~\gamma=0,~~~m=2,~~~\alpha=1,
$$
and assuming the correlations given in Tab. \ref{fleck}.
\begin{table}[!htcb]
\begin{center}
\caption{\footnotesize Correspondence between parameters of (\ref{eq: yield function})-(\ref{gi})
and Deshpande and Fleck (2002) yield functions
---the latter shortened as `DF model'--- to describe
the behaviour of metallic foams.}
\vspace{3 mm}
\label{fleck}
\begin{tabular}{c||c|c|c||c|c||}
\cline{2-6}
  & \multicolumn{3}{c||}{ Model (\ref{eq: yield function})-(\ref{gi})} & \multicolumn{2}{c||}{ DF model} \\  
\cline{2-6}
  & $M$ & $c$ & $p_c$ & $Y$ & $\alpha$ \\ 
\hline
\hline
\multicolumn{1}{||c||}{~}&&&&& \\
\multicolumn{1}{||c||}{$\stackrel{\mbox{DF model}}{\stackrel{~}{Y}, \alpha}$} & $2 \alpha$ &$\ds{\frac{Y}{\alpha}\sqrt{1+\left(\frac{\alpha}{3}\right)^2}}$ & $\ds{\frac{Y}{\alpha}\sqrt{1+\left(\frac{\alpha}{3}\right)^2}}$ & --- & --- \\ 
\multicolumn{1}{||c||}{~}&&&&& \\
\hline
\multicolumn{1}{||c||}{~}&&&&& \\
\multicolumn{1}{||c||}{$\stackrel{\mbox{Model (\ref{eq: yield function})-(\ref{gi})}}{\stackrel{~}{M}, p_c, c}$} & --- & --- & --- & $\ds{\frac{c M}{2\sqrt{1+\left(\frac{M}{6}\right)^2}}}$ & $\ds{\frac{M}{2}}$ \\
\multicolumn{1}{||c||}{~}&&&&& \\
\hline
\hline
\end{tabular}
\end{center}
\end{table}

\begin{figure}[!htcb]
  \begin{center}
      \includegraphics[width= 9 cm]{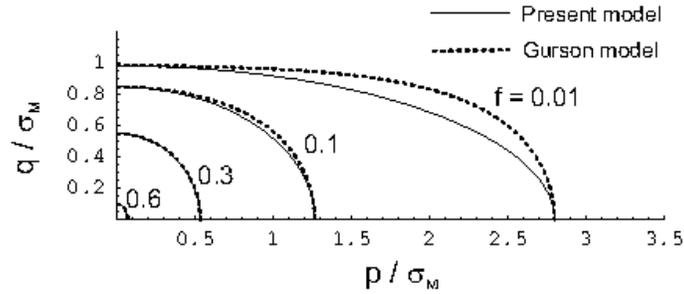}
\caption{\footnotesize Comparison with the Gurson model at different values of void volume fraction f.}
\label{fig10}
  \end{center}
\end{figure}

The proposed function (\ref{eq: yield function})-(\ref{gi}) is also expected to model correctly yielding of 
porous ductile metals. As a demonstration of this, we present in Fig.~\ref{fig10} a comparison 
with the Gurson (1977) model. The Gurson yield function has a circular deviatoric section
so that $\beta=1$ and $\gamma = 0$ in our model, in addition, we select 
\beq
\alpha = 1,~~~m = 2,~~~
p_c = c = \sigma_M \frac{2}{3 q_2} \cosh^{-1}\frac{1+q_3 f^2}{2 f q_1}, ~~~
M = \sigma_M \frac{2}{p_c} \sqrt{1+q_3 f^2 - 2 f q_1},
\eeq
where $f$ is the void volume fraction 
(taking the values \{0.01, 0.1, 0.3, 0.6 \} in Fig.~\ref{fig10}), $\sigma_M$ 
is the equivalent flow stress in the matrix material and
$q_1 = 1.5$, $q_2 = 1$ and $q_3 = q_1^2$ are the parameters introduced by Tvergaard (1981, 1982).
A good agreement between the two models can be appreciated from Fig.~\ref{fig10}, increasing
when the void volume fraction $f$ increases.

\begin{figure}[!htcb]
  \begin{center}
      \includegraphics[width= 10 cm]{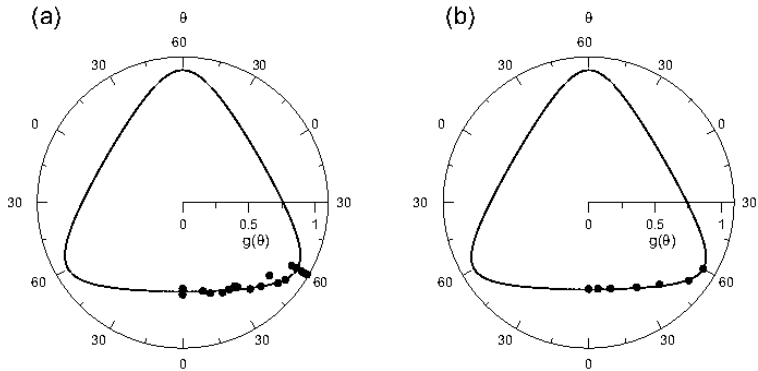}
\caption{\footnotesize Comparison with experimental results relative to deviatoric section
for sandstone (a) and dense sand (b) data taken from (Lade, 1997).}
\label{fig11}
  \end{center}
\end{figure}

As far as the deviatoric section is regarded, we limit to two examples ---reported in Fig. \ref{fig11}---
concerning sandstone and dense sand, where the experimental data have been taken from Lade 
(1997, their Figs. 2 and 9a).

\begin{figure}[!htcb]
  \begin{center}
      \includegraphics[width= 12 cm]{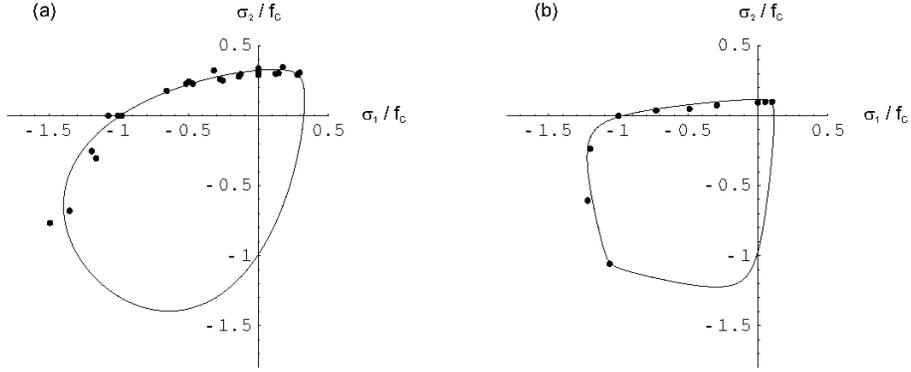}
\caption{\footnotesize Comparison with experimental results on biaxial plane
for cast iron (data taken from Coffin and Schenectady, 1950)(a) and concrete (data taken from Tasuji et al. 1978)(b).}
\label{fig12}
  \end{center}
\end{figure}

Experimental data referred to the biaxial plane $\sigma_3=0$ for grey cast iron and concrete (taken respectively from 
Coffin  and Schenectady, 1950, their Fig. 5 and Tasuji et al. 1978, their Figs. 1 and 2) are reported in Fig. \ref{fig12}.

\section{On convexity of yield function and yield surface}
\lb{conve}

Convexity of the yield function (\ref{eq: yield function})-(\ref{gi})
within the range of parameters (\ref{sevenstones})-(\ref{sevenplus}) was simply stated in the previous Section and 
still needs a proof. This will be given at the end of the present section as an application of a general proposition
relating convexity of yield functions and surfaces that is given below.

\begin{figure}[!htcb]
  \begin{center}
      \includegraphics[width= 8 cm]{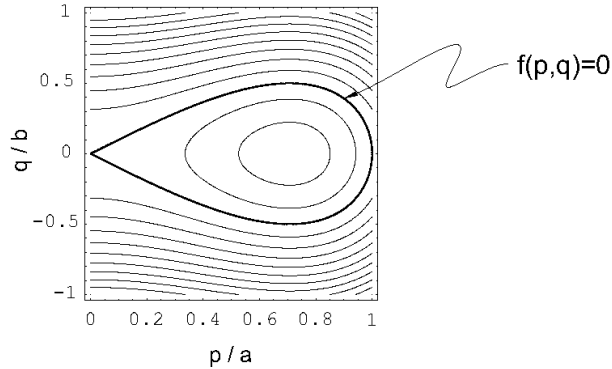}
\caption{\footnotesize Level sets of function (\ref{esempio}).}
\label{fig13}
  \end{center}
\end{figure}

We begin noting that while convexity of yield function implies convexity of the 
corresponding yield surface, the converse is usually false, namely, 
convexity of the level set of a function is unrelated to convexity of the function itself.
As an example, let us consider the non-convex yield function
\beq
\lb{esempio}
f(p,q) = \frac{p^4}{a^4}-\frac{p^2}{a^2} + \frac{q^2}{b^2},~~~0\leq \frac{p}{a} \leq 1,
\eeq
(where $a$ and $b$ are material parameters having the dimension of stress)
which corresponds to a convex yield surface $f(p,q) = 0$, Fig. \ref{fig13}.
After the pioneering work of de Finetti (1949), it became clear that
convexity of {\it every} level set of a function represents its {\it quasi-convexity}, 
a property defining a class of functions much broader than the class of convex functions. 
In more detail, 
let us consider a function $f(\vx): U \longrightarrow \mathbb{R}$, with $U$ being a convex
set, and its level sets
\beq
L_\alpha = \{\vx \in U ~|~ f(\vx) \leq \alpha \},
\eeq
so that: 
\begin{quote}
$f$ is quasi-convex if the level sets $L_\alpha$ are convex for every $\alpha \in \mathbb{R}$. 
\end{quote}
Now, the above definition of quasi-convexity is equivalent (Roberts and Varberg, 1973)
to the definition 
\beq
f[\lambda \vx + (1+\lambda)\vy] \leq \max{\left\{f(\vx), f(\vy)\right\}},~~~
\forall \vx, \vy \in U,~~ \forall \lambda \in [0,1],
\eeq
and, if $f$ is continuous and differentiable, 
to
\beq
\lb{drucker}
f(\vy) \leq f(\vx) ~~\Rightarrow~~\nabla f(\vx) \scalp (\vx-\vy) \geq 0,~~~
\forall \vx, \vy \in U.
\eeq

Convexity of the yield {\it surface} can be either accepted on the basis of experimental 
results, or on some engineering argumentation, such as for instance Drucker's postulate. 
Obviously, a convex yield locus can be expressed as 
a level set of a function, that generally may lack convexity and, even, quasi-convexity. 
For example, the level sets of function 
(\ref{esempio}) are given in Fig. \ref{fig13}. 
It can be observed that while $f(p,q)=0$ may perfectly serve as a (convex) 
yield surface, the corresponding 
yield function even lacks quasi-convexity\footnote{As noted by Franchi et al. (1990), 
definition (\ref{drucker}) 
is very similar to Drucker's postulate. However, Drucker's postulate merely prescribes 
the so-called normality rule of plastic flow and convexity of yield surface (Drucker, 1956, 1964). 
Quasi-convexity
becomes a consequence of Drucker's postulate only in the special case ---considered by 
Franchi et al. (1990)--- in which convexity of yield surface implies convexity of 
all level sets of the corresponding function.}. It is true that, in principle, a convex yield function
can always be found to represent a convex yield surface,
but to find this in a reasonably simple form may be an hard task. In other words, 
a number of yield functions that were formulated as an interpolation of experimental results still need 
a proof of convexity, even in cases where the corresponding yield locus is convex.
The propositions that will be given below set some basis to provide these proofs.

\subsection{A general result for a class of yield functions}

The yield function (\ref{eq: yield function})-(\ref{gi}) presented in the previous section 
may be viewed as an element of
a family of models specified by the generic form (\ref{eq: yield function}). This family
includes, among others, the models by Gudheus (1973), 
Argyris et al. (1974), Willam and Warnke (1975), Eekelen (1980), Lin and Bazant (1986), 
Bardet (1990), Ehlers (1995), and Men\'{e}trey and Willam (1995) and Christensen (1997)
and Christensen et al. (2002).

A general result is provided below showing that for the range of material parameters
for which the Haigh-Westergaard 
representation of a yield surface (\ref{eq: yield surface 2})
is convex, the function is also convex. 

\vspace*{3mm}
\noindent {\it Proposition 1}:
Convexity of the yield {\it function} (\ref{eq: yield function}) is equivalent to convexity 
of the me\-ri\-dian and deviatoric sections of the corresponding yield {\it surface} (\ref{eq: yield surface 2})
in the Haigh-Westergaard representation. 
In symbols:
\beq
\lb{propo}
{\rm convexity~of~} F(\stress) = f(p) + \frac{q}{g(\theta)}
 \Longleftrightarrow f''\geq0, ~~~\& ~~~ g^2+2g'\,^2-gg'' \geq 0,
\eeq
where $g(\theta)$ is a positive function.
\vspace*{3mm}

{\it Proof.} It is a well-known theorem of convex analysis (Ekeland and Temam, 1976) that the sum of two 
convex functions is also a convex function. Since $q$, $p$ and $\theta$ are independent parameters, failure
of convexity of $f(p)$ or $q/g(\theta)$ implies failure of convexity of $F(\stress)$ and therefore
convexity of both $f(p)$ and $q/g(\theta)$ are necessary and sufficient conditions for convexity of $F(\stress)$.

Now, let us first analyze $f(p)$. The fact that convexity of $f(p)$ as a function of 
$\stress$ is equivalent to convexity of the 
meridian section follows from linearity of the trace operator, in view of the fact that $p = -\tr \stress /3$.

Second, the fact that convexity of $q/g(\theta)$ as function of $\stress$ 
is equivalent to the convexity of the deviatoric section follows from the 3 lemmas listed below.

\hspace{140mm}$\Box$

\vspace*{3mm}
\noindent {\it Lemma 1} (Hill, 1968): 
Convexity of an isotropic function of a symmetric (stress) tensor $\stress$ is equivalent to convexity of the
corresponding function of the principal (stress) values $\sigma_i$ ($i=1,2,3$).
In symbols, given:
\beq
\phi(\stress) = \tilde{\phi}(\sigma_1, \sigma_2, \sigma_3),
\eeq
then:
\beq
\Delta \frac{\partial \phi}{\partial \stress} \scalp \Delta\stress \geq 0 \Longleftrightarrow
\sum_{i=1}^3 \Delta \frac{\partial \tilde{\phi}}{\partial \sigma_i} \, \Delta\sigma_i  \geq 0,
\eeq
where $\Delta$ denotes an ordered difference in the variables, so that, denoting with $A$ and $B$
two points in the tensor space
\beq
\Delta \frac{\partial \phi}{\partial \stress} \scalp \Delta\stress = 
\left(\left. \frac{\partial \phi}{\partial \stress}\right|_{\stress^A}
-\left. \frac{\partial \phi}{\partial \stress}\right|_{\stress^B}\right)\scalp(\stress^A-\stress^B) \, ,
\eeq
\beq
 \Delta \frac{\partial \tilde{\phi}}{\partial \sigma_i} \, \Delta\sigma_i = 
\left(\left. \frac{\partial \tilde{\phi}}{\partial \sigma_i}\right|_{\{\sigma_1^A,\, \sigma_2^A,\, \sigma_3^A\}}
-\left.\frac{\partial \tilde{\phi}}{\partial \sigma_i}\right|_{\{\sigma_1^B,\, \sigma_2^B,\, \sigma_3^B\}}\right)
(\sigma_i^A-\sigma_i^B)\, ,
~~~i=1,2,3.
\eeq
\vspace*{3mm}

{\it Proof.} That convexity of $\phi$ implies convexity of $\tilde{\phi}$ is self-evident. The converse is 
not trivial and a proof was given by Hill (1968) with reference to an elastic strain energy function.
The proof, omitted here for brevity, was later obtained also by Yang (1980) with explicit reference to 
a yield function.

\hspace{140mm}$\Box$

\vspace*{3mm}
\noindent {\it Lemma 2}: 
Given a generic isotropic function $\phi$ of the stress that can be expressed as
\beq
\phi(\sigma_1, \sigma_2, \sigma_3) = \tilde{\phi}(S_1,S_2),
\eeq
where $S_1$ and $S_2$ are two of the principal components of deviatoric stress, i.e.
\beq
\lb{devia}
S_1= \frac{1}{3} \left(2\sigma_1-\sigma_2-\sigma_3\right),~~~
S_2= \frac{1}{3} \left(-\sigma_1+2\sigma_2-\sigma_3\right),
\eeq
convexity of $\phi(\sigma_1, \sigma_2, \sigma_3)$ is equivalent to convexity of $\tilde{\phi}(S_1,S_2)$.
\vspace*{3mm}

{\it Proof.} The proof follows immediately from the observation that the relation (\ref{devia}) between 
$\{S_1, S_2\}$ and $\{\sigma_1, \sigma_2, \sigma_3\}$ is linear.

\hspace{140mm}$\Box$

\vspace*{3mm}
\noindent {\it Lemma 3}: 
Convexity of 
\beq
\lb{roteta}
\frac{q}{g(\theta)}
\eeq
as a function of $S_1, S_2$ is equivalent to the convexity of the deviatoric section in the Haigh-Westergaard space:
\beq
\lb{curvatura}
g^2 + 2g'\,^2 -g g'' \geq 0.
\eeq
\vspace*{3mm}

{\it Proof.} The Hessian of (\ref{roteta}) is
\beq
\lb{iniz}
\begin{array}{ll}
\ds{ \frac{\partial^2 q/g(\theta)}{\partial S_i \partial S_j}}&= \ds{\frac{1}{g^3}} \left[
\ds{g^2 \frac{\partial^2 q}{\partial S_i\partial S_j}  + q (2g'\,^2 -g g'') 
\frac{\partial \theta}{\partial S_i} \frac{\partial \theta}{\partial S_j}}
\right. \\ [5 mm]
& \left. \ds{-g\, g' \left( \frac{\partial q}{\partial S_i} \frac{\partial \theta}{\partial S_j}
+\frac{\partial q}{\partial S_j} \frac{\partial \theta}{\partial S_i}
+q \frac{\partial^2 \theta}{\partial S_i  \partial S_j}
\right)}
\right],
\end{array}
\eeq
where $i$ and $j$ range between 1 and 2 and all functions $q$ and $\theta$ are to be understood
as functions of $S_1$ and $S_2$ only. Derivatives of $q$ may be easily calculated to be
\beq
\frac{\partial q}{\partial S_i} = 2 S_i - (-1)^i m_i, 
~~~\frac{\partial^2 q}{\partial S_i \partial S_j} = \frac{27}{4\, q^3} m_i \, m_j,
\eeq
where indices are not summed and vector ${\bf m}$ has the components
\beq
\{{\bf m} \}= \{S_2, -S_1\}.
\eeq
The derivatives of $\theta$ can be performed through $\cos 3\theta$, eqn. 
(\ref{invarianti2})$_1$, noting that 
\beq
\begin{array}{l}
\ds{\frac{\partial \theta}{\partial S_i}  = \frac{-1}{3 \sin 3\theta}\frac{\partial \cos 3\theta}{\partial S_i}}, 
\\ [5 mm]
\ds{\frac{\partial^2 \theta}{\partial S_i \partial S_j}  = \frac{-1}{3 \sin 3\theta} \left(
\frac{\cos 3\theta}{\sin^2 3\theta}
\frac{\partial \cos 3\theta}{\partial S_i}
\frac{\partial \cos 3\theta}{\partial S_j}
+
\frac{\partial^2 \cos 3\theta}{\partial S_i \partial S_j} \right)},
\end{array}
\eeq
so that 
\beq
\lb{pizza}
\begin{array}{l}
\ds{\frac{\partial q}{\partial S_i} \frac{\partial \theta}{\partial S_j}
+\frac{\partial q}{\partial S_j} \frac{\partial \theta}{\partial S_i}
+q \frac{\partial^2 \theta}{\partial S_i  \partial S_j}} \\ [5 mm]
= 
\ds{ \frac{-1}{\sin 3\theta}
\left[
\frac{\partial^2 q \cos 3\theta}{\partial S_i \partial S_j}-\cos 3\theta \frac{\partial^2q}{\partial S_i \partial S_j}
+q\frac{\cos 3\theta}{\sin^2 3\theta} 
\frac{\partial\cos 3\theta}{\partial S_i} \frac{\partial \cos 3\theta}{\partial S_j}
\right]},
\end{array}
\eeq
where
\beq
\lb{treteta}
\frac{\partial \cos 3\theta}{\partial S_i} = \frac{9 \sqrt{3}\, \sin 3 \theta}{2 \, q^2} 
~ m_i,~~~
\frac{\partial^2 q\cos 3\theta}{\partial S_i \partial S_j} = -27^2 \frac{J_3}{q^6} m_i \, m_j.
\eeq
A substitution of (\ref{treteta}) into (\ref{pizza}) yields
\beq
\ds{\frac{\partial q}{\partial S_i} \frac{\partial \theta}{\partial S_j}
+\frac{\partial q}{\partial S_j} \frac{\partial \theta}{\partial S_i}
+q \frac{\partial^2 \theta}{\partial S_i  \partial S_j}}
= 0,
\eeq
so that we may conclude that the Hessian (\ref{iniz}) can be written as
\beq
\lb{hessian}
 \frac{\partial^2 q/g(\theta)}{\partial S_i \partial S_j} 
 = \frac{27}{4}
\frac{\left( g^2 + 2g'\,^2 -g g'' \right)}{q^3 g^3} m_i \, m_j.
\eeq
Positive semi-definiteness of the Hessian (\ref{hessian}) is condition (\ref{curvatura}), which, in turn, 
represents non-negativeness of the curvature (and thus convexity) of deviatoric section.

\hspace{140mm}$\Box$

\subsection{Applications of Proposition 1}

The scope of this section is on one hand to prove the convexity of function (\ref{eq: yield function})-(\ref{gi})
within the range (\ref{sevenstones})-(\ref{sevenplus}) of material parameters,
on the other hand to show that 
Proposition 1 is constructive, in the sense that can be used  to invent convex yield functions.
Let us begin with the first issue.

\subsubsection{The proposed yield function (\ref{eq: yield function})-(\ref{gi})}

First, we show that $f(p)$, eqn (\ref{eq: yield function}), is a 
convex function of $p$ (so that the meridian section is convex) 
and, second, that the 
deviatioric section described by $g(\theta)$, eqn (\ref{gi}), is convex 
for the range of material parameters listed in (\ref{sevenstones})-(\ref{sevenplus}). Therefore,
as a conclusion from Proposition 1, function $F(\stress)$ results to be convex.

A well-known result of
convex analysis (Ekelan and Temam, 1976) states that function $f(p)$ is convex if and only if
the restriction to its effective domain (i.e. $\Phi \in [0, 1]$) is convex.
Moreover, the function $\Phi$ appearing in (\ref{fiegi})$_1$ is a
linear function of $p$ so that convexity of $f(p)$ can be inferred from convexity of
the corresponding function, say $\tilde{f}$, of $\Phi$.
Introducing for simplicity the function
\beq
\lb{acca}
h(\Phi) = \left(\Phi-\Phi^m\right)\left[2(1-\alpha)\Phi+\alpha\right],
\eeq
the convexity of function $\tilde{f}(\Phi)$ reduces to the condition
\beq
\lb{con}
\left[h^{\,\prime}(\Phi)\right]^2
-2\,h^{\,\prime\prime}(\Phi)h(\Phi) \geq 0,
\eeq
where
\beq\lb{derivativesh}
\barr{l}
h^{\,\prime}(\Phi)=\left(1-m\Phi^{m-1}\right)
\left[2(1-\alpha)\Phi+\alpha\right]
+2(1-\alpha)\left(\Phi-\Phi^m\right), \\ [3 mm]
h^{\,\prime\prime}(\Phi)=-m(m-1)\Phi^{m-2}
\left[2(1-\alpha)\Phi+\alpha\right]
+4(1-\alpha)\left(1-m\Phi^{m-1}\right).
\earr
\eeq
Fulfillment of eqn (\ref{con}) can be now easily proven considering the inequality
\beq
h^{\,\prime\prime}(\Phi)\leq4(1-\alpha)\left(1-m\Phi^{m-1}\right),
~~~\forall\ \Phi\in[0,1].
\eeq

It remains now to show convexity of $q/g(\theta)$. To this purpose, Proposition 1 can 
be employed, through substitution of (\ref{gi}) 
into the convexity condition eqn (\ref{curvatura}), thus yielding
\beq
\lb{dumarron}
\frac{1}{g(\theta)} +
\frac{3\gamma\cos 3 \theta}{\sqrt{1-\gamma^{2}\cos^{2} 3\theta}}
\sin\left[\beta\frac{\pi}{6}-\frac{1}{3}\cos^{-1}(\gamma\cos 3\theta)\right] \geq 0,
\eeq
where $\theta \in [0, \pi/3]$ and $g(\theta)$ is given by eqn (\ref{gi}).
For values of $\gamma$ belonging to the interval specified in (\ref{sevenstones})$_7$,
condition (\ref{dumarron}) can be transformed into
\beq
\sin\left(\frac{\beta\pi}{6}-\frac{4}{3}x\right)+2\sin\left(\frac{\beta\pi}{6}+\frac{2}{3}x\right) \geq 0,
\eeq
with $x \in [\cos^{-1}\gamma,~ \pi - \cos^{-1}\gamma]$ and then into
\beq
\frac{-1+2 \cos z +2 \cos^2 z}{2\,\sin z (1-\cos z)}\sin \beta\frac{\pi}{6} + \cos \beta\frac{\pi}{6} \geq 0,
\eeq
with $z \in [2/3 \cos^{-1}\gamma,~ 2/3(\pi - \cos^{-1}\gamma)]$, an inequality that can be shown to be verified within the interval of 
$\beta$ specified in (\ref{sevenplus}) and thus also within its subinterval (\ref{sevenstones})$_6$.

\subsubsection{Generating convex yield functions}

Proposition 1 can be easily employed to build convex yield functions within the class described 
by eqn (\ref{eq: yield function}).
The simplest possibility is
to maintain $f(p)$ in the form (\ref{effedip}) and change the deviatoric function eqn (\ref{gi}).
As a first proposal, we can introduce the following function
\beq
\lb{gi2}
g(\theta)=
\left[1+\beta\left(1+\cos3\HWt\right)\right]^{-1/n},
\eeq
instead of (\ref{gi}). This describes a smooth deviatoric section approaching (without reaching) the triangular 
(Rankine) shape when parameters 
$n > 0$ and $\beta \geq 0$ are varied. The yield function is convex within the range of parameters reported 
in Tab.~\ref{pb} (see Appendix B for a proof).
\begin{table}[!htcb]
\begin{center}
\caption{\footnotesize Conditions for the convexity of deviatoric yield function (\ref{gi2}).}
\vspace{3 mm}
\label{pb}
\begin{tabular}{||c|c||}
\hline\hline 
 & \\ [-2 mm]
$0< n \leq 11/3$   &$n\geq 11/3$            \\  [3 mm]
\hline\hline
                                  &              \\
$\ds{\beta \leq \frac{n}{9-2n}}$&  $\ds{\beta\leq\left(-1+\sqrt{1+  
\frac{9(n-2)^2}{n^2(4n-13)}} \, \right)^{-1}}$
                  \\
 & \\
\hline\hline
\end{tabular}
\end{center}
\end{table}
The yield function defined by eqns (\ref{effedip}) and (\ref{gi2}) does not possess the extreme deformability
of (\ref{effedip}) and (\ref{gi}) and does not admit 
Mohr-Coulomb and Tresca as limits, but results in a simple expression.
\begin{figure}[!htcb]
  \begin{center}
      \includegraphics[width= 10 cm]{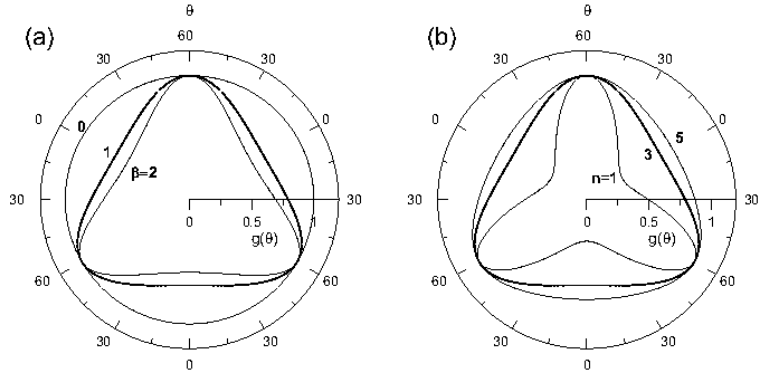}
\caption{\footnotesize Deviatoric section (\ref{gi2}): 
effects related to the variation of $\beta$ (a) and $n$ (b).}
\label{fig14}
  \end{center}
\end{figure}
The performance of the deviatoric shape of the yield surface 
is analyzed in Fig.~\ref{fig14}, where the solid lines correspond to the limit of convexity, $\beta=1$ and $n=3$. 
The curves reported in Fig. \ref{fig3} (a) are relative to the values of $\beta=0, 1, 2$,
whereas for Fig. \ref{fig3} (b) $n$ takes the values $\{1, 3, 5 \}$.

A limitation of the yield surface described by eqns (\ref{effedip}) and (\ref{gi2}) 
is that the deviatoric section cannot be stretched 
until the Rankine limit. This can be easily emended 
assuming for $g(\theta)$ our expression (\ref{gi}) or 
that proposed by Willam and Warnke (1975) (see also Men\'etrey and Willam, 1995)
\beq
\lb{willam}
g(\theta) = 
\frac{2 (1-e^2) \cos\theta +(2 e -1) \left[ 4 (1-e^2) \cos^2\theta + 5 e^2 -4 e\right]^{0.5}}
{4(1-e^2) \cos^2\theta + (2 e -1)^2} ,
\eeq
where $e \in ]0.5,1]$ is a material parameter, yielding in the limit $e \longrightarrow 0.5$ the
Rankine criterion and the von Mises criterion when $e=1$.

It is already known that the deviatoric section of the yield surface corresponding to eqn
(\ref{willam}) remains convex for any 
value of the parameter $e$ ranging within the interval $]0.5, 1]$, so that ---from Proposition 1--- the
function (\ref{eq: yield function}) equipped with the definition (\ref{willam}) of the 
function $g(\theta)$ is also convex. 

As a final example, we can employ function $g(\theta)$ defined by the expression
proposed by Gudheus (1973) and Argyris et al. (1974) 
\beq
\lb{pa}
g(\theta) = \frac{2k}{1 + k + (1-k) \cos{3\theta}},
\eeq
where $k \in ]0.777, 1]$ is a material parameter. 

Otherwise, we can act on the meridian function. For instance, 
we can modify a Drucker-Prager criterion ---which again fits in the framework described by
eqn (\ref{eq: yield function})--- obtaining a non-circular deviatoric section 
described by eqn (\ref{gi}) 
\beq
F(\stress) = - \Gamma \left(p +c\right)+ q
\cos{\left[ \beta \frac{\pi}{6} - \frac{1}{3}\cos^{-1}\left(\gamma \cos{3 \theta}\right)\right]},
\eeq
where $c$ is the yield strength under isotropic tension and $\Gamma$ is a material parameter, or by eqn (\ref{gi2})
\beq
\lb{db}
F(\stress) = - \Gamma \left(p +c\right) + q [1 + \beta (1+\cos 3 \theta)]^{1/n},
\eeq 
or by the Gudheus/Argyris condition (\ref{pa})
\beq
\lb{da}
F(\stress) = - \Gamma \left(p +c\right) + \frac{q}{2k}\left({1 + k + (1-k) \cos{3\theta}}\right).
\eeq
It may be noted that the yield criterion (\ref{da}) has been employed by Laroussi et al. (2002) 
to describe the behaviour of foams.
In all the above cases, Proposition 1 ensures that for the range of parameters in which the 
Haigh-Westergaard representation of the yield surface is convex, the yield function is also convex.

\subsection{A note on the behaviour of concrete and a generalization of Proposition 1}

In the modelling of concrete there is some experimental evidence that the deviatoric
section starts close to the Rankine limit for low hydrostatic stress component and tends to approach 
a circle, when confinement increases.
This effect has been described by 
Ottosen (1977) through a model which
does not fit the general framework specified by eqn (\ref{eq: yield function}) and can be written in our
notation in the 
form
\beq
\lb{ottos}
F(\stress) = A q^2 + B\frac{q}{g(\theta)} + C - p ,
\eeq
where $A >0$, $B \geq 0$ and $C\leq 0$ are constants and $g(\theta)$ is in the form (\ref{gi})
with $\beta=0$. The criterion is therefore defined by four parameters.

The above-expression (\ref{ottos}) of the yield function suggests the following generalization of
Proposition 1:

\vspace*{3mm}
\noindent {\it Proposition 2}:
Convexity of the yield function 
\beq
\lb{generaliz}
F(\stress) = A q^2+ B \frac{q}{g(\theta)} + f(p),  
\eeq
where $A$ and $B$ are positive constants, is equivalent to
\beq
\lb{conv}
f''\geq0, ~~~\& ~~~ g^2+2g'\,^2-gg'' \geq 0,
\eeq
\vspace*{3mm}
which in turn is equivalent to the convexity of the surface 
\beq
B \frac{q}{g(\theta)} + f(p) = 0,
\eeq
in the Haigh-Westergaard stress space.

{\it Proof.} Let us begin assuming that (\ref{conv}) holds true. In this 
condition Proposition 1 ensures that $f(p)$ and 
$q/g(\theta)$ are convex 
functions of $\stress$, so that (\ref{generaliz}) results the sum of three convex functions 
and its convexity follows.
Vice-versa, failure of convexity of $f(p)$ immediately implies failure of convexity of (\ref{generaliz})
since $p$ is independent of $\theta$ and $q$. Finally, let us assume that condition (\ref{conv})$_2$ 
is violated, for a certain value, say $\tilde{\theta}$, of $\theta$. The Hessian of 
\beq
A q^2+ B q/g(\theta), 
\eeq
as a function of two components of deviatoric stress $S_1$ and $S_2$, is given by eqn. (\ref{hessian}) summed to 
a constant and positive definite matrix
\beq
\lb{hessian2}
3A \left[
\begin{array}{cc}
2 & 1 \\ [5 mm]
1 & 2 
\end{array}
\right]
+B\frac{27}{4}
\frac{\left( g^2 + 2g'\,^2 -g g'' \right)}{q^3 g^3}
\left[
\begin{array}{cc}
S_2^2 & -S_1S_2 \\ [5 mm]
-S_1S_2 & S^2_1 
\end{array} \right] .
\eeq
Considering now the Haigh-Westergaard representation, 
it is easy to understand that we can keep $\theta = \tilde{\theta}$
fixed 
and change $S_1$ and consequently $S_2$ so that $S_1/S_2$ remains constant. 
In this situation, while $g$ and its derivatives remain fixed, the quantity 
\beq
\frac{S_1^2}{q^3},
\eeq
in matrix (\ref{hessian2}) 
tends to $+\infty$ when $S_1$ tends to zero. Therefore, violation of (\ref{conv})$_2$ cannot be compensated by a
constant term and function (\ref{generaliz}) is not convex.

\hspace{140mm}$\Box$

Proposition 2 provides the conditions for convexity of the Ottosen criterion. Moreover, the same proposition
allows us to generalize our yield function (\ref{eq: yield function})-(\ref{gi}) 
adding a $q^2$ term as in the Ottosen criterion. This leads immediately to
\beq
F(\stress) = A \, q^2 + B \, q 
\cos{\left[ \beta \frac{\pi}{6} - \frac{1}{3}\cos^{-1}\left(\gamma \cos{3 \theta}\right)\right]}
+ f(p),
\eeq
where $f(p)$ is given by eqn (\ref{effedip}).

\section{Conclusions}

In the modelling of the inelastic behaviour of several materials,
the knowledge of a smooth, convex yield surface approaching known-criteria and possessing an extreme shape variation 
to fit experimental results may be of undoubted utility. In particular, the fact that a yield function can 
continuously describe a transition between yield surfaces typical of different materials is of fundamental
importance in modelling the de-cohesion due to damage of rock-like materials and the increase in cohesion during
forming of powders.
In the present
paper, such a yield function has been proposed, which is shown
to be capable of an accurate description of the behaviour of a broad
class of materials including soils, concrete, rocks, powders, metallic foams, porous materials, and polymers.
Moreover, in order to analyze convexity of our function, we have provided certain general results, 
holding for a broad class of yield conditions, which permit to infer
convexity of the yield function from convexity of the yield surface.

\vspace*{3mm}
\noindent
{\sl Acknowledgments }

\noindent
The authors are grateful to Dr. Alessandro Gajo (University of Trento)
for many useful discussions and suggestions.
Financial support from University of
Trento, Trento, Italy is gratefully acknowledged.

\newpage

{
\singlespace

}

\clearpage
\setcounter{equation}{0}
\renewcommand{\theequation}{{A}.\arabic{equation}}
\begin{center}
{\bf APPENDIX A. Values of material parameters employed to fit experimental results in Sec. \ref{fitting}}\\
\end{center}

The values of the material parameters of the proposed yield function (\ref{eq: yield function})-(\ref{gi})
employed to model the experimental data reported
in Figs. \ref{fig5}-\ref{fig9} and Figs.~\ref{fig11}-\ref{fig12} are listed in Table~\ref{values} below.

\noindent
\begin{table}[!htcb]
\begin{center}
\caption{\footnotesize Values of material parameters employed to fit experimental results in 
Figs.~\ref{fig5}-\ref{fig9} and Figs.~\ref{fig11}-\ref{fig12}.}
\vspace{3 mm}
\label{values}
\begin{tabular}{c||c|c|c|c|c|c|c||}
\cline{2-8}
 & $M$ & $p_c$ & $c$ & $m$ & $\alpha$ & $\beta$ & $\gamma$ \\
\hline\hline
\multicolumn{1}{||c||}{Fig. \ref{fig5}(a)} & $0.5$ & $0.961$ MPa & $0$ & $2.6$ & $0.1$ & $0$ & $0.9999$ \\
\hline
\multicolumn{1}{||c||}{Fig. \ref{fig5}(b)} & $0.48$ & --- & $0$ & $5$ & $0.2$ & $0$ & $0.66$ \\
\hline
\multicolumn{1}{||c||}{Fig. \ref{fig6}(a)} & $1.1$ & --- & $0$ & $3.2$ & $1.9$ & --- & --- \\
\hline
\multicolumn{1}{||c||}{Fig. \ref{fig6}(b)} & $0.76$ & --- & $0$ & $3.4$ & $1.85$ & --- & --- \\
\hline
\multicolumn{1}{||c||}{Fig. \ref{fig6}(c)} & $0.94$ & --- & $0$ & $1.8$ & $0.8$ & --- & --- \\
\hline
\multicolumn{1}{||c||}{Fig. \ref{fig6}(d)} & $0.64$ & --- & $0$ & $3$ & $1.5$ & --- & --- \\
\hline
\multicolumn{1}{||c||}{Fig. \ref{fig7}(a)} & $0.82$ & $p_c=100~f_c$ & $c=0.1~f_c$ & $2$ & $0.05$ & --- & --- \\
\hline
\multicolumn{1}{||c||}{Fig. \ref{fig7}(b)} & $0.76$ & $p_c=100~f_c$ & $c=0.2~f_c$ & $2$ & $0.026$ & --- & --- \\
\hline
\multicolumn{1}{||c||}{Fig. \ref{fig8}(a)} & $1.18$ & $p_c =50~f_c$ & $c =0.04~f_c$ & $2$ & $0.04$ & --- & --- \\
\hline
\multicolumn{1}{||c||}{Fig. \ref{fig8}(b)} & $0.61$ & $p_c =30~f_c$ & $c =0.02~f_c$ & $2$ & $0.3$ & --- & --- \\
\hline
\multicolumn{1}{||c||}{Fig. \ref{fig9}(a)} & $0.40$ & $1000$ MPa & $110$ MPa& $2$ & $0.5$ & --- & --- \\
\hline
\multicolumn{1}{||c||}{Fig. \ref{fig9}(b)} & $0.51$ & $800$ MPa& $80$ MPa& $2$ & $0.62$ & --- & --- \\
\hline
\multicolumn{1}{||c||}{Fig. \ref{fig11}(a)} & --- & --- & --- & --- & --- & $0$ & $0.843$\\
\hline
\multicolumn{1}{||c||}{Fig. \ref{fig11}(b)} & --- & --- & --- & --- & --- & $0$ & $0.862$\\
\hline
\multicolumn{1}{||c||}{Fig. \ref{fig12}(a)} & $0.78$ & $p_c/f_c=5.7$ & $c/f_c=0.3$ & $2$ & $0.1$ & $0$ & $0.6$\\
\hline
\multicolumn{1}{||c||}{Fig. \ref{fig12}(b)} & $0.86$ & $p_c/f_c=3.5 \times 10^7$ & $c/f_c=0.08$ & $1 \times 10^6$ & $1 \times -10^8$ & $0.12$ & $0.98$\\
\hline\hline
\end{tabular}
\end{center}
\end{table}

\clearpage
\setcounter{equation}{0}
\renewcommand{\theequation}{{B}.\arabic{equation}}
\begin{center}
{\bf APPENDIX B. Covexity of function (\ref{gi2})}\\
\end{center}

\noindent
We show that the deviatoric section described by eqn (\ref{gi2}) is convex,
within the range of parameters reported in 
Tab. \ref{pb}. To this purpose, a substitution of eqn (\ref{gi2}) into 
condition (\ref{curvatura}) yields
\beq
\lb{eq: convexity condition 2}
a \cos^{2}(3\HWt) 
+b \cos(3\HWt) + c \geq0 ,~~~\HWt\in\left[0, \pi/3\right],
\eeq
where the coefficients $a$, $b$ and $c$ are:
\beqar
a & = & \beta^{2}\left(n^{2}-9\right),      \\
b & = & \beta n(1+\beta)(9-2n),             \\
c & = & n^{2}(1+\beta)^{2}+9\beta^{2}(1-n).
\eeqar
Since condition (\ref{eq: convexity condition 2}) is bounded by a parabola, it suffices
to analyze the position of its vertex to obtain the values of parameters reported in Tab. 
\ref{pb}.



\end{document}